\documentclass{my-article}

\articletype{Research Article}

\usepackage[normalem]{ulem} 
\usepackage{siunitx} 
\usepackage{lineno}
\usepackage{pgfplots}
\usetikzlibrary{spy}
\usepackage[makeroom]{cancel}

\pgfplotsset{compat=1.18} 

\newcommand{\fref}[1]{Fig.~\ref{#1}}
\newcommand{\rb}{$^{87}$Rb} 

\begin{document}

\title{All-fiber, near-infrared, laser system at 780~nm for atom cooling}

\author{
Matteo Marchesini\authormark{1}\authormark{*}, 
Michelangelo Dondi\authormark{1},
Leonardo Rossi\authormark{2},   
Gabriele Bolognini\authormark{2},
Marco Prevedelli\authormark{1}, and
Francesco Minardi\authormark{1}
}

\address{\authormark{1} Dipartimento di Fisica e Astronomia, Alma Mater Studiorum Università di Bologna, 40127 Bologna, Italy\\
\authormark{2} Consiglio Nazionale delle Ricerche, Istituto per la Microelettronica e i Microsistemi, 40129 Bologna, Italy}

\email{\authormark{*}m.marchesini@unibo.it} 


\begin{abstract} 
One of the prominent platforms for quantum technologies, cold atoms require reliable laser systems. We present the design, implementation, and characterization of a simple, compact, and economical laser system at 780~nm, entirely based on fiber components. Two semiconductor lasers at 1560~nm are amplified in a single Erbium-doped fiber amplifier and frequency-doubled in a periodically-poled lithium niobate crystal. We characterize the amplitude noise and the linewidths of the lasers, as well as the SHG efficiency. With a rms relative amplitude noise of $ 3\times 10^{-4}$ at 1~s and linewidths below 1~MHz, our system is suitable for cooling and trapping of Rb atoms.
\end{abstract}

\section{Introduction}
Quantum technologies based on cold atoms call for integrated, reliable and robust experimental setups, in order to lead to commercial devices. Notable examples of quantum sensors are the atom interferometers \cite{Kasevich1991} used as accelerometers \cite{Peters2001,Menoret2018,Stray2022}, already marketed as gravimeteres \cite{companies} and gyroscopes \cite{Gustavson1997,Chen2019} where integration is paramount. 

Among the key elements of all these setups and devices are the laser sources needed for atom cooling and manipulation. The workhorse of atom cooling is Rubidium (Rb), for which the cooling D2 transition $5\, ^{2}S_{1/2} \rightarrow 5\, ^2 P_{3/2}$ has a wavelength of 780~nm. For Rb-based experiments, the widely used approach is to employ semiconductor lasers, suitably narrowed in frequency, e.g. by means of an extended-cavity \cite{Ricci1995, Baillard_2006}, and amplified with other semiconductor devices - such as tapered amplifiers - to make up for their typically low power. These sources  provide radiation with power of the order of several hundreds mW and spectral width of the order of 1~MHz.

Over the last years an alternative solution has been developed, taking advantage of the fact that 780~nm light can be obtained by second harmonic generation of 1560~nm infrared radiation \cite{Thompson_2003}. Since 1560~nm lies within the C-band of fiber telecommunications, a panoply of commercial fibers devices (sources, splitters and combiners, polarizers, amplifiers, detectors \dots) is readily available, with well-characterized performances. Thus, the alternative solution is to process the laser beams in the infrared domain, and perform second-harmonic generation of 780~nm light at the very end. Several works have shown the advantages of this solution: San\'e {\it et al.} amplify in fiber a single 1560nm frequency and generate 11W at 780nm with a in-air crystal \cite{Sane_2012}; Dong {\it et al.}, with a similar approach, obtain more than 20W \cite{Dong_2019}. Other works show architectures using 2 in-series SHG stages: Theron {\it et al.} efficiently convert two separate seed lasers frequency-frequency separated by hundred GHz \cite{Theron_2017}, while Kim {\it et al.} use an amplifier and SHG crystal in series for each seed laser to achieve 40W power output \cite{Kim_2020}. All these systems present many fibered components, but have the common drawback of performing the frequency doubling in free-space, and/or using more than one amplifying/doubling component. Some fully-fibered commercial options are available, with outputs in the order of 1~W and tunable in a frequency range $\lesssim$1~GHz only \cite{muquans:ILS}.

In this paper, we describe the experimental realization of a simple, compact and relatively economical laser system for Rb atom cooling, comprising two sources at 1560~nm, differing in frequency by approximately 3~GHz, combined and amplified in the same Erbium-doped fiber amplifier (EDFA) and then frequency-doubled in a single non-linear crystal with fiber input and output. The system is fully fibered from source to output, and is stabilized through an external 780~nm reference laser, allowing a frequency tunability in a wide range (0.1$\div$6.5~GHz). 

\section{Architecture description}

\begin{figure}[t]
\includegraphics[width=\textwidth]{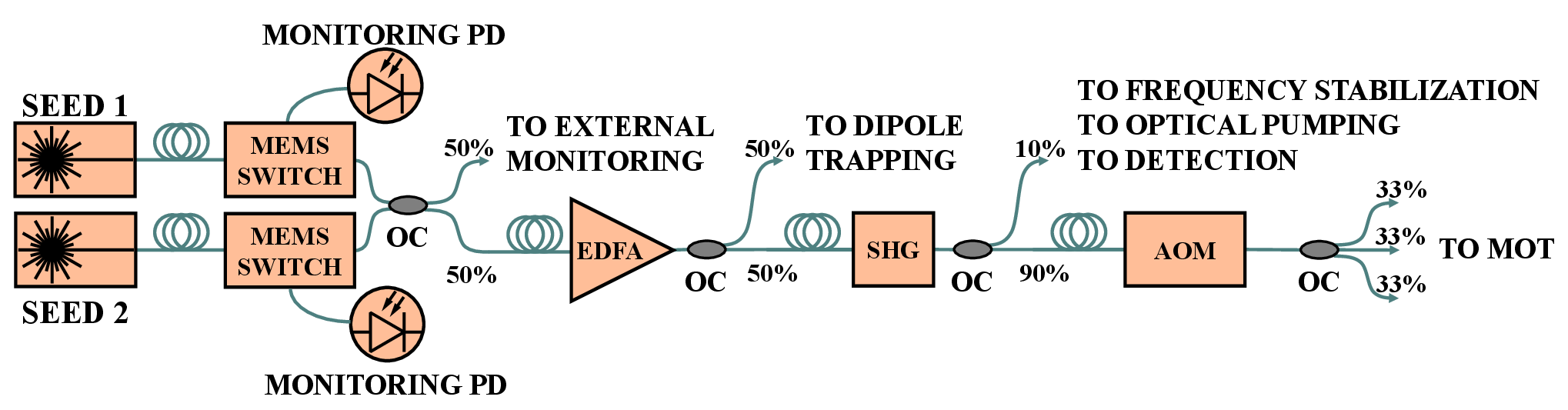}
\caption{Sketch of the system architecture:  EDFA Erbium-doped-fiber-amplifier, SHG second-harmonic generation crystal, AOM acousto-optic modulator, OC optical combiner, PIN photodiode.}
\label{fig:setup}
\end{figure}

We describe the general architecture of the laser system, depicted in \fref{fig:setup} and designed to meet the requirements for laser cooling and trapping \rb\ atoms in a magneto-optical trap. Two distinct laser frequencies are needed since the ground electronic level $5 ^2 S_{1/2}$ is split by hyperfine interactions in two sublevels, $|F=1\rangle$ and $ | F=2\rangle$, separated in energy by $\Delta E_{\rm HFS}/h = 6.83468$~GHz, that are both populated by collisions and by optical pumping. The two lasers, commonly referred to as "cooler" and "repumper", are usually imbalanced in power by one order of magnitude because most of the laser power is needed for the cooling transition $(F=2\rightarrow F'=3)$, while less power is sufficient for the repumping one $(F=2\rightarrow F'=2)$. For both lasers, and most notably for the cooler, the frequency agility - {\it i.e.} the possibility to perform frequency jumps and/or rapid sweeps of tens of MHz - is highly advantageous, because different stages of the laser cooling/trapping protocol require frequency (as well as amplitude) variations in the ms timescale.

We start with two semiconductor lasers at 1560~nm (Eblana EP-1550-0-NLW), with a nominal power of 5~mW -- at the maximum current of 240~mA -- and nominal Lorentzian linewidth of 150~kHz; actually the measured powers of the two lasers, that increase linearly with the injection current above the threshold value $I_{th} \simeq 33$ mA, are larger than the nominal value, being 5.7 and 7.5~mW already at injection current of 200~mA.

Both laser output fibers are connected to a MEMS switch (OF-LinkMEMS-1x2-PM-1560-M2-P-09-10-FA) and then to a 2 × 2 power combiner (OF-Link FPMC-1560-22-50-PM-L-10-FA-B). As we wish to amplify both lasers with the same polarization, we cannot use a 2 × 1 polarization power combiner, whereby two beams are combined into the same output port with orthogonal polarization. One output port of the combiner is relayed to an Erbium-doped fiber amplifier (EDFA), with a 30dB gain and a maximum output power of 5 W (Keopsys CEFA-C-PB-HP), while the other output port can be used for external monitoring of the seeds.
Power monitoring of the lasers is performed with a couple of photodiodes connected to the secondary exits of the MEMS.

The EDFA output is frequency-doubled by a second-harmonic-generation (SHG), periodically poled, lithium niobate (PPLN) crystal, with fiber input and output (NTT WH-0780-000-F-B-C). Since the maximum input power allowed in the PPLN crystal is only 500 mW, the EDFA output is split by a $1\times 2$, 50/50, polarization beam splitter/combiner (OF-Link HPMFC-1560-12-50-PM-L-10-FA-B-05)  and half power at 1560 nm is used for far off-resonant trapping of atoms. 

The SHG converts more than 50\% of the injected power, generating nearly 300~mW of light at 780~nm that are divided in a $1\times 2$, 10/90, polarization beam splitter (OF-Link FPMC-780-12-10-PM-L-10-FA-B). Thus, 90\% of 780~nm light is used for the three retroreflected beams of the magneto-optical trap, while the remaining 10\% fraction goes for frequency stabilization, optical pumping and detection. The 90\% port is relayed to an acousto-optic modulator (Aerodiode 780AOM-1), operating at 100 MHz, that is inserted for quick amplitude control of the MOT beams. Finally, the AOM output is fed to a $1\times3$ power splitter (OF-Link FPMC-780-13-333-PM-L-10-FA-B) to generate three beams of equal power for the magneto-optical trap. The 10\% port is combined, in a fiber power combiner, with a 780~nm reference laser (not shown in \fref{fig:setup}) locked to Rb saturated absorption, allowing a beat-note frequency stabilization with a tuning range from 0.1 to 6.5~GHz, that covers both the "cooler" and "repumper" transitions described above.

\section{Amplitude and frequency noise at 1560 nm}

Here we report the amplitude and frequency noise characterization of our laser sources, starting from the seed lasers at 1560~nm. As a prelude, we present the noise measurement of our home-built current sources used to drive the seed lasers. 

\subsection{Current noise of laser driver}

\begin{figure}[t]
\includegraphics[width=\textwidth]{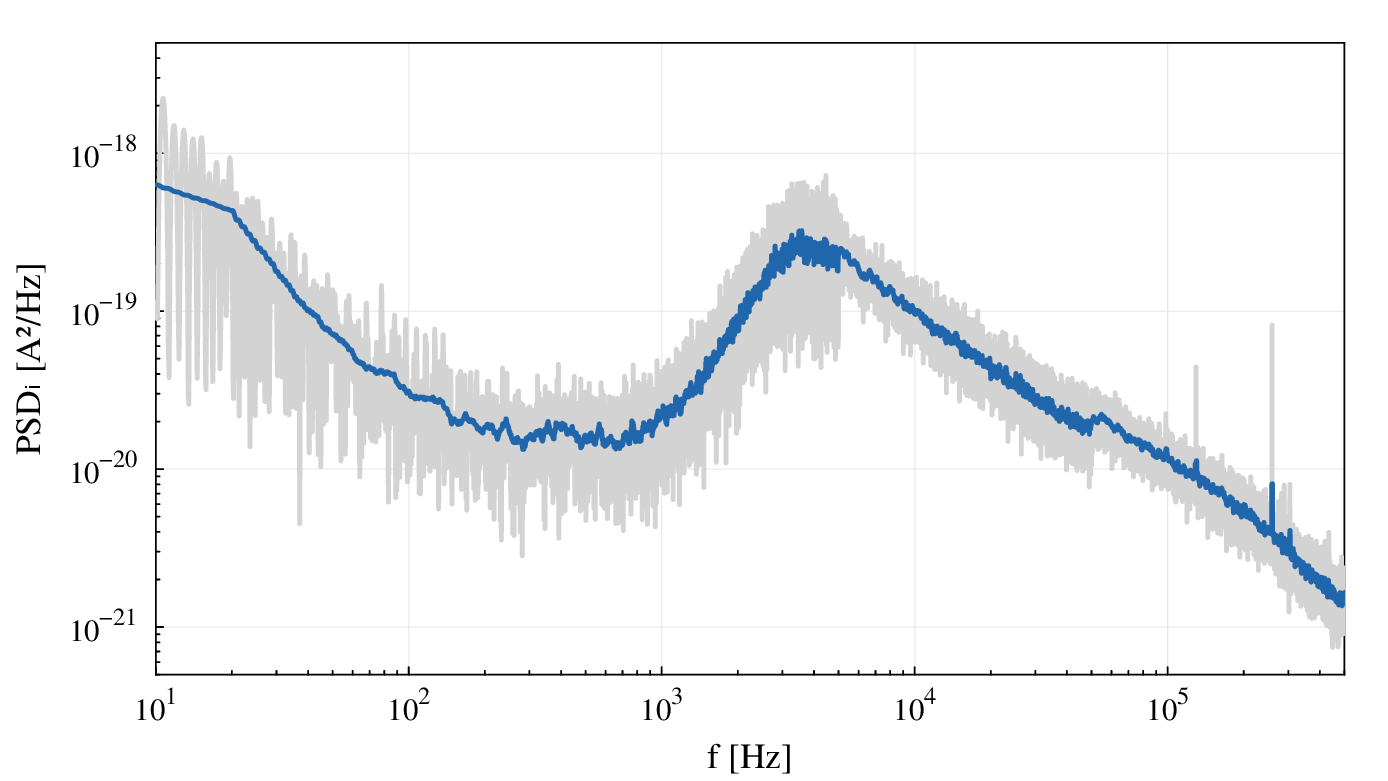}
\caption{Power spectral density (PSD) of the current driving the seed lasers. Grey line: PSD calculated as FFT of scope trace; blue line: 256-points running average, smoothing the trace.}
\label{fig:LDDnoise}
\end{figure}

The current drivers are based on a well known circuit described
originally in \cite{Libbrecht1993} and upgraded to be digitally
controlled in \cite{Erickson2008}. They can provide up to 250 mA and
have a modulation input with a peak deviation of $\pm 1$~mA with a
bandwidth of about 200~kHz and are completely remote--controlled via
an I2C bus.

The current noise power spectral density (PSD) of the drivers has been
measured at 10~mA using a low noise transimpedance amplifier in the
10~Hz--500~kHz bandwidth by acquiring data in the time domain and
computing their FFT with a digital oscilloscope. The results are shown in \fref{fig:LDDnoise}. The current noise
integrated in the above mentioned bandwidth amounts to about 74~nA
rms. In terms of laser amplitude noise, since the power increases with the driving current at a rate of \SI{35}{\micro\watt}/mA, the RMS current noise contributes to the RMS power noise by 2.6 nW, {\it i.e.} 0.5 ppm of our working power.

In general, the current noise is more relevant for the laser frequency noise. After converting the current PSD to the frequency PSD using the
measured tuning coefficient of 500~MHz/mA, the contribution of the current noise to laser linewidth has been evaluated by  the $\beta$--separation line
method, described in \cite{DiDomenico2010}: it amounts to 50~kHz, which is nearly negligible given the nominal linewidth of 150~kHz.

\subsection{Seed laser RIN}
\label{subsec:seed_RIN}

\begin{figure}[t]
\includegraphics[width=\textwidth]{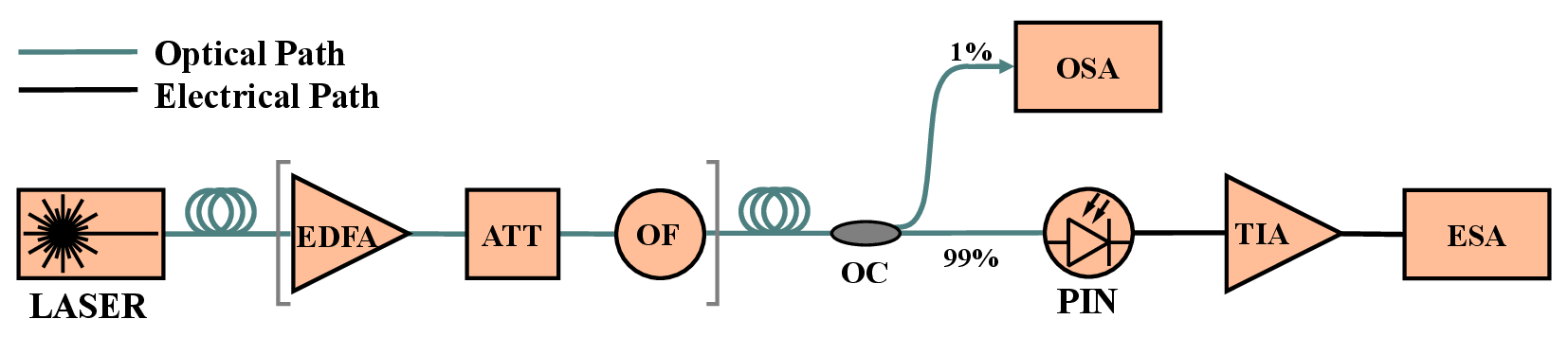}
\caption{Setup for intensity noise measurement:  ATT attenuator, OF optical filter, PIN photodiode, TIA transimpedence amplifier, OSA/ESA optical/electronic spectrum analyzer. The parts in bracket are present only for the measurements of the amplified radiation.
}
\label{fig:RINMEAS}
\end{figure}

\begin{figure}[t]
\centering
\includegraphics[width=\textwidth]{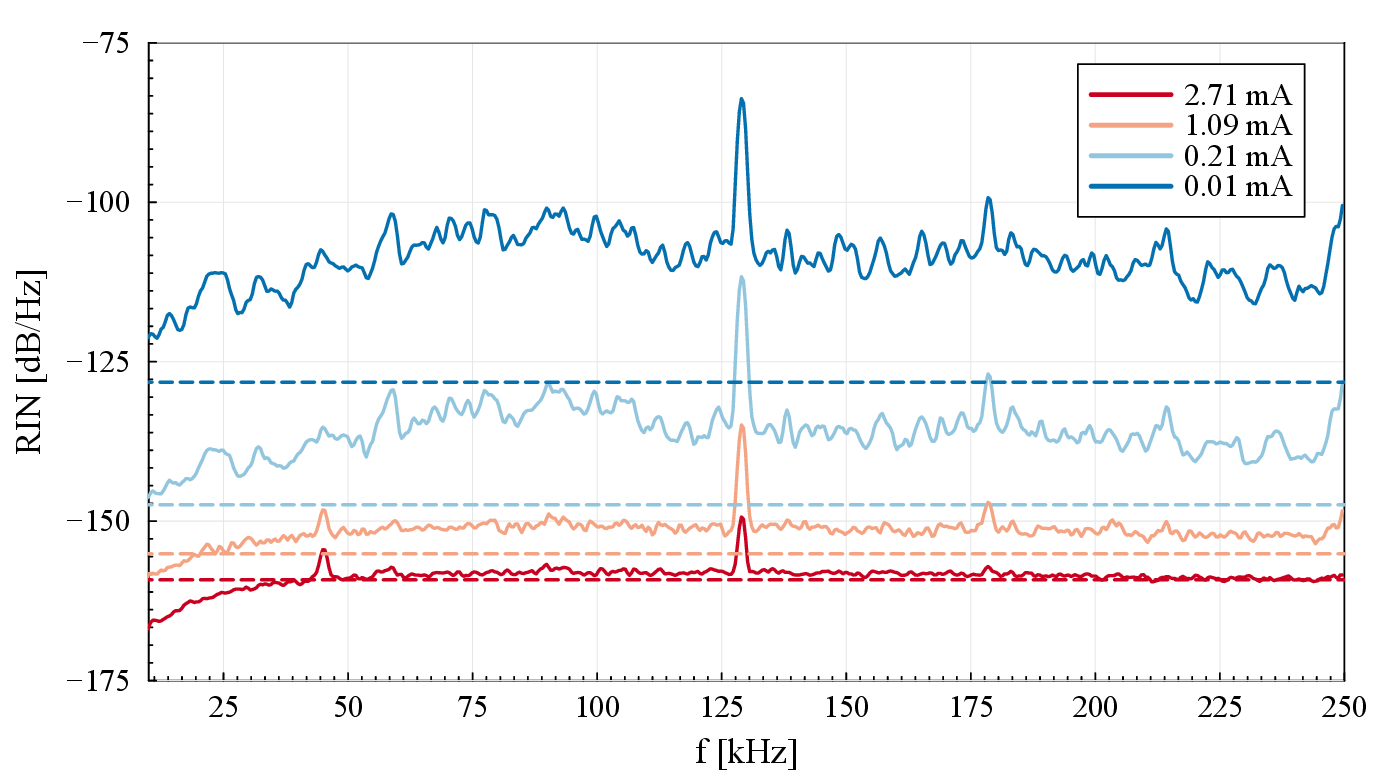}
\caption{
Calculated RIN based on the ESA traces for different values of the dc photocurrent, hence power; dashed lines indicate the RIN baseline due to shot-noise and Johnson thermal noise.}
\label{fig:YE4140}
\end{figure}

To measure the seed laser amplitude noise we use an InGaAs detector with FC/APC input receptacle (Thorlabs DET010CFC); the photodiode is inversely biased with an internal 5V voltage regulator, powered by a 12V battery, and it has a 1\si{\kilo\ohm} resistor connected in series to the output. When followed by a transimpedance amplifier (TIA), the internal resistor is irrelevant for current-to-voltage conversion and merely limits the maximum photocurrent. Indeed, the detector output is fed to a TIA (Femto DLPCA-200) with variable current-to-voltage gain ($G$) and bandwidth, whose output is analyzed with a spectrum analyzer (Rigol  DSA875) or a digital oscilloscope (Rigol DS2302A). The measurement setup is shown in \fref{fig:RINMEAS}, omitting the part within square brackets. The electrical spectrum analyzer (ESA) nominal range is 9~kHz to 1.5~GHz; at high frequency we are limited by the TIA bandwidth, below 9 kHz we Fourier-transform the digital oscilloscope trace.

Intensity noise fluctuations are quantified through relative intensity noise (RIN), which is defined as the Fourier transform of the optical power auto-correlation function $C_P (t,\tau)$:

\begin{align}
    \mathrm{RIN} (f ) &=\int_{-\infty}^{+\infty} C_P(\tau ) e^{-i 2 \pi f \tau} d\tau \\
    C_P (t,\tau) &= \frac{\langle \delta P_{\mathrm{opt}} (t) \delta P_{\mathrm{opt}} (t+\tau) \rangle}{\bar P_{\mathrm{opt}}^2}
\end{align}
where  $\bar P_{\mathrm{opt}}$ 
is the average of optical power, $\delta P_{\mathrm{opt}}(t) = P_{\mathrm{opt}}(t) - \bar P_{\mathrm{opt}}$ represents time-dependent power fluctuations and $\langle \rangle$ denotes the time average. 

The reading of the ESA is related to the above defined RIN as follows:

\begin{align}
\label{eq:RIN}
    S(f) Z_\mathrm{in} &= \frac{G^2 I_\mathrm{dc}^2 B}{Z_\mathrm{in} } \mathrm{RIN} (f) \\
         \Rightarrow \mathrm{RIN  [dB/Hz]} &= 10 \log_{10}[S(f) Z_\mathrm{in}/G^2 I_\mathrm{dc}^2 B ] 
\end{align}
where $I_\mathrm{dc}$ the dc photocurrent output by the detector, $G$ the transimpedance gain, and $Z_\mathrm{in}$ = 50 \si{\ohm} the ESA input impedance.  

In \fref{fig:YE4140}, we show the corresponding RIN for one of the two seed lasers, namely Eblana YE4140, which is designed as Cooler. Different traces correspond to different values of the laser power, from slightly above the lasing threshold to nearly maximum, and correspondingly different values of the dc photocurrent $I_\mathrm{dc}$. We have verified that the other seed laser, Eblana YE4142, designed as Repumper, has similar traces.

When photocurrent fluctuations are solely due to shot-noise, {\it i.e.} $I_{\mathrm{rms}} (f,B)^2 = \sigma^2_{I,shot} = 2e I_\mathrm{dc} B$, then $\mathrm{RIN} = 2eB/I_\mathrm{dc}$. In addition, Johnson thermal noise on the internal resistor $R_L$ is to be taken into account, contributing to $I_{\mathrm{rms}}^2$ with $\sigma^2_{I,J} = 4 (k_B T/R_L) B$. 
In our case the photodiode has a responsivity very close to 1 A/W, corresponding to a quantum efficiency of 0.8: at the largest supply current and power, the Johnson noise is negligible with respect to shot-noise, which amounts to 27 pA/$\sqrt{\si{\hertz}}$, and nearly accounts for the measured RIN \cite{TIA_input_noise}.

\subsection{Seed laser frequency noise}
\label{subsec:freq_noise1560}

\begin{figure}[t]
\begin{center}
\includegraphics[width=\textwidth]{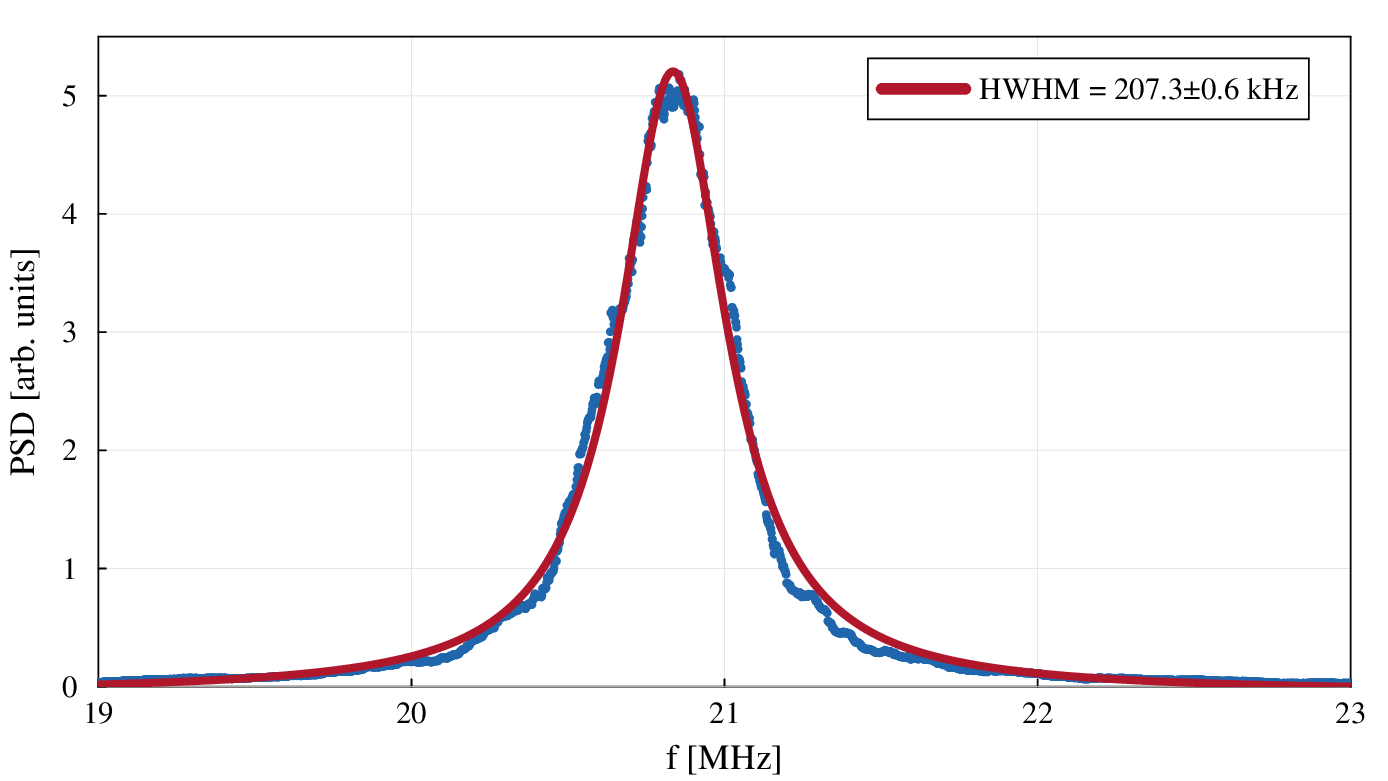}  
\end{center}
\caption{
PSD of the beat note signal generated by the seed lasers, calculated as the FFT of the digital scope time traces. A moving average over 101 point is applied upon the $y$~vector of 3601 points.}
\label{fig:beat1560}
\end{figure}

To characterize the frequency noise of our lasers, we perform a heterodyne measurement placing an InGaAs photodiode with integrated transimpedance amplifier (New Focus 1811-FC) on the unused output port of the 2 x 2 power combiner. A 20~dB power attenuator is necessary for the input light not to exceed the saturation power of \SI{50}{\micro\watt}. The detector has a nominal conversion of 40~kV/W and a bandwidth of 125~MHz. We adjust the temperature and driving current of the two laser to have their frequency difference at approximately 19~MHz, well within the accepted bandwidth.  

We observe the beatnote of the two sources by computing the FFT of time-domain traces acquired on a digital scope. We find this method more convenient than the use of an ESA because it allows a shorter interrogation time, shielding from the low frequency instabilities. We acquire traces of total length equal to \SI{240}{\micro\second} and compute the PSD of the photocurrent \cite{Steck:2023}: 

\begin{align}
    PSD_i(\omega) &\propto \int_{-\infty}^\infty R(\tau) e^{i\omega \tau} d\tau 
    \\
    R(\tau ) &\equiv \langle E^*(t)E^*(t+\tau)E(t+\tau)E(t) \rangle \nonumber\\
    &= (|A_1|^2 + |A_2|^2 )^2 + |A_1|^2|A_2|^2  e^{-\frac12\langle \varphi_{12}^2 (\tau) \rangle}
    \left( e^{i \Delta_{12}\tau } + \mathrm{c.c.} \right) 
    \label{eq:PSD} 
\end{align}
with $E(t) = \sum_{j=1,2} A_j e^{i\omega_j t} e^{i\phi_j(t)}$, $\Delta_{12}~\equiv \omega_1 -\omega_2$ is the frequency difference and $\varphi_{12} (\tau)~\equiv~\phi_{12}(t+\tau) - \phi_{12}(t)$ and $\phi_{12}(t) = \phi_1(t) -\phi_2(t)$; clearly,

\begin{align}
    \langle \varphi_{12}^2 (\tau) \rangle &= 2 C_{12}(0) - 2 C_{12}(\tau) \\
    C_{12} (\tau) &\equiv \langle \phi_{12}(t) \phi_{12}(t+\tau) \rangle
\end{align}
and, since for two independent sources the mutual phase correlations vanish,

\begin{align}
\label{eq:c12}
    C_{12} (\tau) &= C_1(\tau) + C_2(\tau) .
\end{align}

The full-width at half-maximum (FWHM) of the lasers, hence of their beatnote, depends on the type of phase noise \cite{DiDomenico2010}.  It can be shown that, in case of white frequency noise, $C_i(\tau) = C_i(0) - \gamma_i |\tau| \, (i=1,2)$, and $2\gamma_i$ are the FWHM of the Lorentzian-broadened PSD of the laser electric fields. In this case, Eq.~\ref{eq:c12} shows that also the photocurrent PSD displays a Lorentzian peak at the beatnote frequency, with a FWHM equal to the sum of the lasers FWHM. From \fref{fig:beat1560} we see that a Lorenzian fit is well suited to the detected beatnote photocurrent PSD and we obtain a FWHM of $\approx$207~kHz for each laser, slightly larger than the nominal value of 150~kHz.

\subsection{Amplitude noise of EDFA output}

\begin{figure}[t]
\centering
\begin{tikzpicture}
\node at (0,0) {
    \includegraphics[width=0.475\textwidth]{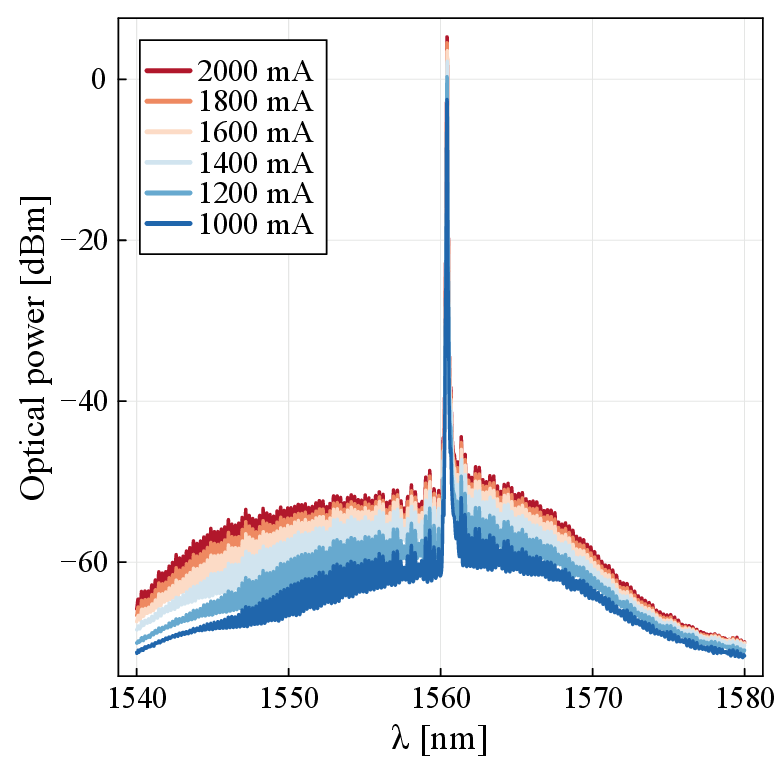}};
\node at (6.75,0) {
    \includegraphics[width=0.475\textwidth]{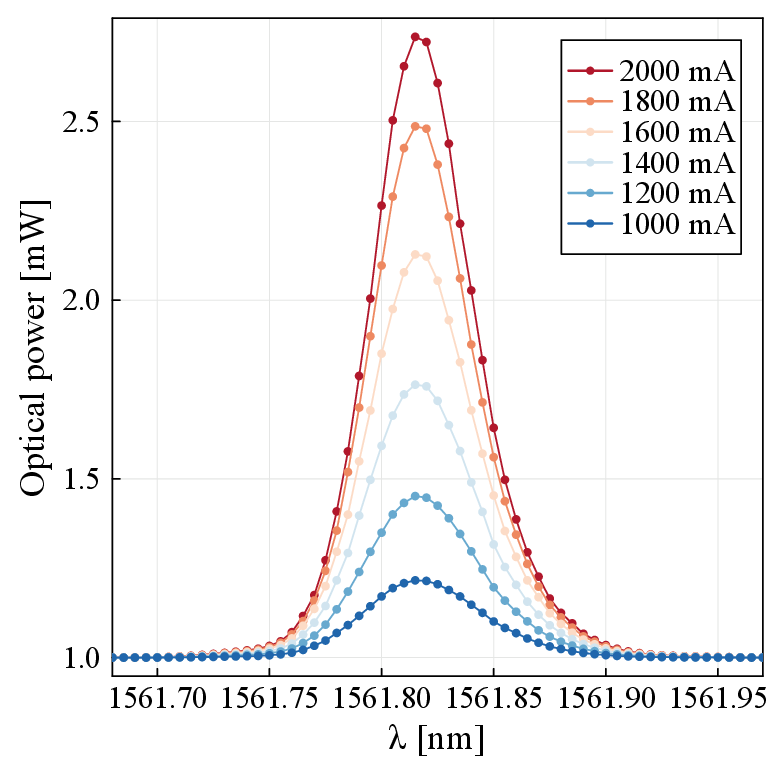}};
\node at (-3,2.75) {\LARGE{(a)}};
\node at (3.75,2.75) {\LARGE{(b)}};
\end{tikzpicture}
\caption{(a) Spectrum of the EDFA output signal, with an input power of 2~mW from the seed laser. (b) Zoomed-in spectrum of the amplified signal, in linear y-scale. }
\label{fig:OSA}
\end{figure}

\begin{figure}[t]
\centering
\begin{tikzpicture}
\node at (0,0) {    
\includegraphics[width=0.475\textwidth]{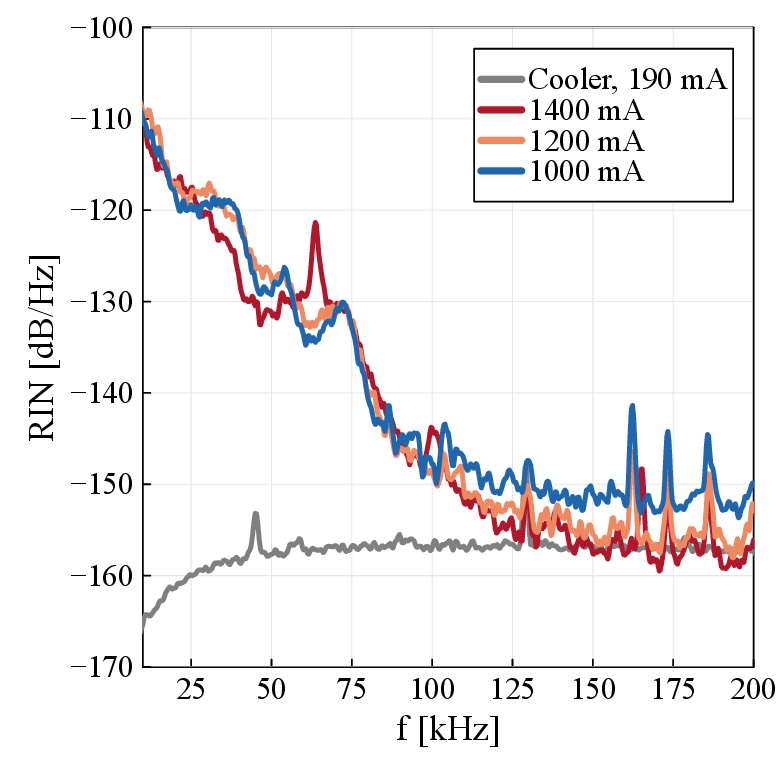}
};
\node at (6.75,0) {    
\includegraphics[width=0.475\textwidth]{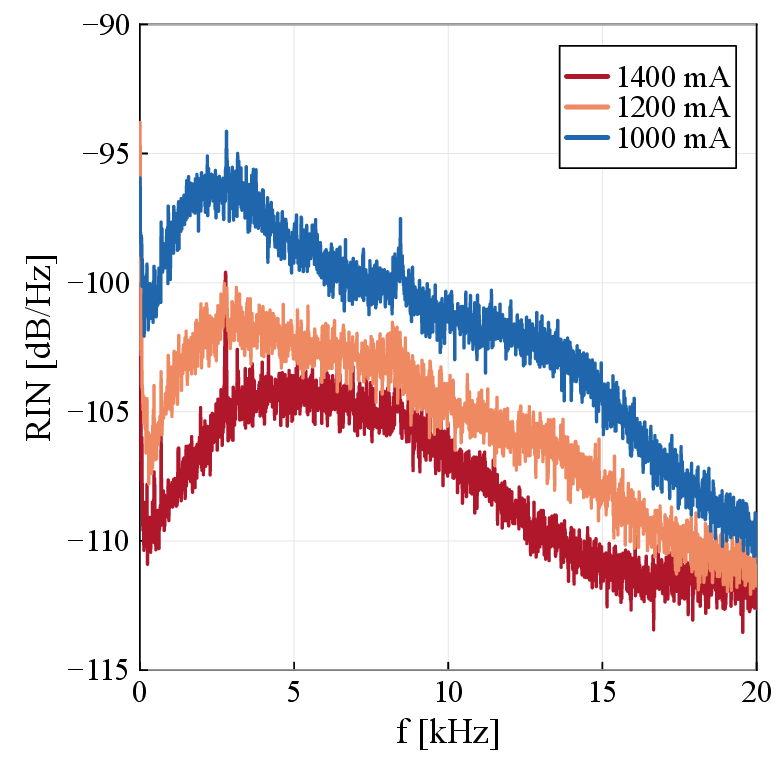}
};
\node at (-3,2.75) {\LARGE{(a)}};
\node at (3.75, 2.75) {\LARGE{(b)}};
\end{tikzpicture}
\caption{(a) RIN of EDFA output at different EDFA pump currents; for comparison, the RIN of the Cooler seed laser (YE4140) reported above is also shown; (b) RIN at low frequencies, as obtained from FFT of digital oscilloscope, below the ESA cut-off (10 kHz).}
\label{fig:EDFA-RIN}
\end{figure}

To assess the performance of the laser system, we have also characterized the amplitude noise of the amplified radiation delivered by the EDFA. The measurement setup is also shown in \fref{fig:RINMEAS}, where the elements in square brackets are now included: we used a 20~dB attenuator (Thorlabs FA20T-APC, ATT in figure) to reduce the EDFA output power below the photodetector saturation value, {\it i.e.} photocurrent < 5mA. In addition, we inserted a tunable optical filter (Micron Optics FFP-TF, OF in figure) based on a Fabry-Pérot etalon, to filter out unwanted broadband amplified spontaneous emission (ASE), which is caused by the optical pumping within the Erbium doped fiber: indeed, a fraction of excited Er atoms decay to their ground level by spontaneous emission, instead of stimulated emission, resulting in a broadband incoherent light spectrum that adds to the amplified seed light \cite{Desurvire}. To evaluate the overall contribution of the ASE to the EDFA emission, the optical filter was removed, and the attenuated EDFA output was coupled into an optical spectrum analyzer (OSA, model EXFO FTB-5240S/BP). The resulting spectra are shown in \fref{fig:OSA}. In particular, \fref{fig:OSA}~(a) reports the spectra of amplified laser source for different values of EDFA pump current; the observed ASE background noise has a broadband spectrum approximately centered around the signal wavelength of 1560~nm. At this specific wavelength, the ratio between the power of the amplified signal and the one of the ASE (signal-to-noise ratio of ASE - SNRASE, 0.1~nm resolution bandwidth), is 55~dB at all pump current values. 

Integrating the spectra obtained in \fref{fig:OSA} over a range from 1540 to 1580~nm, and taking into account the fact that the measurements were taken with a resolution bandwidth of 0.1~nm, we obtain a fractional ASE power of approximately 0.04\%, indicating a small impact of ASE to the amplified light in all employed pump current conditions.

The EDFA output was also characterized in terms of RIN. The measurement set-up was  analogous to the one performed for the semiconductor laser, using the 99\% branch in \fref{fig:RINMEAS} and the photodetector and the process is identical to the one described in Subsection \ref{subsec:seed_RIN}. As was done there, the RIN is extracted from the output of the ESA, normalized with respect to average power, gain and measurement bandwidth using Eq.~\ref{eq:RIN}.

\fref{fig:EDFA-RIN} displays the RIN frequency spectra (in dB/Hz) of the EDFA output light vs frequency for different pump current values (red, purple and blue curves for current values of 1400, 1200 and 1000~mA respectively). These curves are compared with the RIN spectrum of the corresponding seed light (same as blue "2.71 mA" trace in 
\fref{fig:YE4140}).
The results in \fref{fig:EDFA-RIN} show that EDFA amplification adds significantly, i.e. up to 50 dB, 
to the intensity noise below $\sim  150$~kHz, while at higher frequencies the RIN of the amplified light and the RIN of to the seed light are similar. Such an increase in RIN after EDFA amplification  is expected: 
actually the intensity noise of the amplified light is affected by a variety of different noise sources, especially at lower frequencies, 
with a significant contribution of the intensity noise of EDFA pump lasers 
\cite{Putra2017}. Indeed, similar RIN profiles as the one shown in \fref{fig:EDFA-RIN} were also found in earlier works with EDFA-amplified laser sources \cite{Cruz2015}.

It is to note that the frequency range of RIN measurements above  is limited by the TIA bandwidth (200~kHz).
Although RIN values of amplified and non-amplified sources do not differ significantly beyond 150~kHz, as a further verification of the low-RIN behaviour beyond 200~kHz frequency, RIN measurements were repeated, up to 10 MHz, with a New Focus 1811-FC InGaAs photodiode, featuring an integrated TIA and a larger bandwidth (125~MHz). For frequencies beyond 200~kHz, it was found that the RIN values for both the seed laser and the EDFA output were similar, thus confirming that the EDFA only adds intensity noise for frequencies below 150~kHz.

\section{Second-harmonic generation}

\begin{figure}[t]
\centering
\begin{tikzpicture}
\node at (0,0) {    
\includegraphics[width=0.475\textwidth]{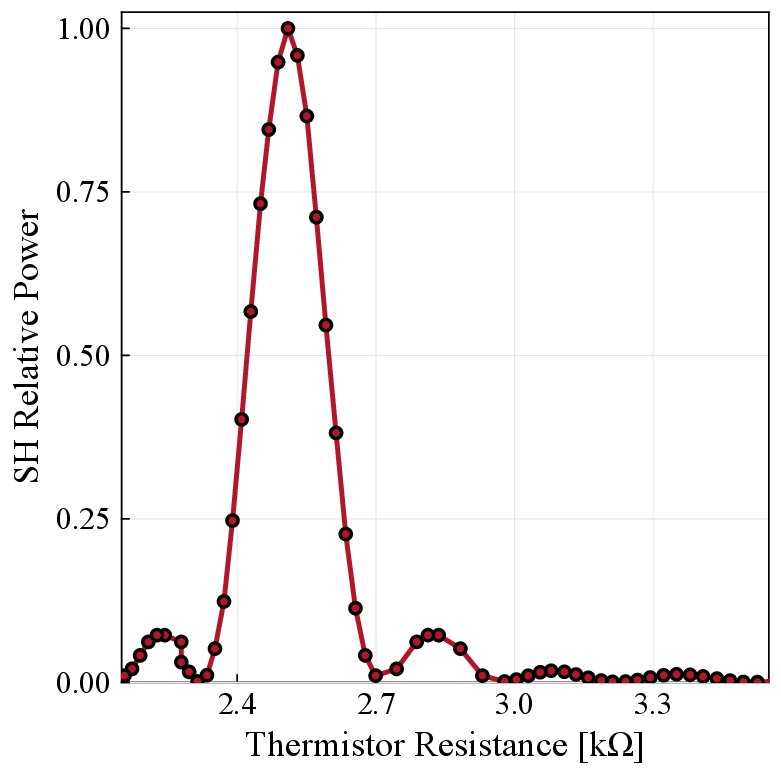}};
\node at (6.75,0) {
\includegraphics[width=0.475\textwidth]{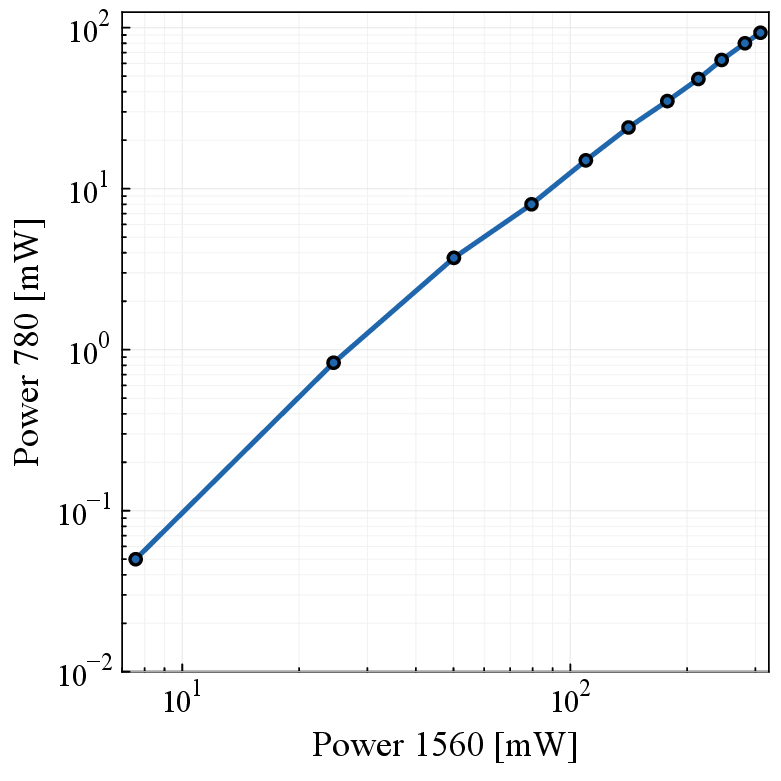}};
\node at (-3.2,2.75) {\LARGE{(a)}};
\node at (3.75,2.75) {\LARGE{(b)}};
\end{tikzpicture}

\caption{(a) Normalized SHG power as a function of the measured resistance of the internal thermistor. The resistance-to-temperature conversion is provided by the manufacturer, according to the empirical equation $R = R_0 \exp(B/T - B/T_0)$, with $R_0=10$~k$\Omega$, $B=3456$~K, $T_0=317$~K; (b) 780~nm output power as a function of the 1560~nm input power, showing the non-linear conversion of the SHG, at the temperature of maximum conversion.}
\label{fig:SHGeff}
\end{figure}

The light output of the amplifier is injected to a 1 x 2 fibered power splitter, and one of the two ouput port is connected to the input fiber of the second-harmonic generator, {\it i.e.} the periodically poled lithium niobate crystal with fiber input and output. The device is rated to accept a maximum injection power of 500~mW at the fundamental wavelength, which is at 1560~nm in our case.

Periodically-poling (PP) is greatly beneficial to ease the ``phase-matching'' constraints required for optimal wavelength conversion \cite{Yariv}. In a PP crystal, the ``quasi phase-matching'' condition is achieved when

\begin{align}
   \left| \frac{2\pi}{\lambda_2} (n_1 - n_2) - m \frac{2 \pi}{\Lambda}  \right| \ll \frac1{L_c}  
\end{align}
where $\lambda_2$ is the vacuum wavelength of the second harmonic, $\Lambda$ the poling period, $m$ an integer number (usually $\pm 1$), $L_c$ the crystal length, and $(n_1, n_2)$ are the refraction indexes for the fundamental and second harmonic waves, respectively.

Since $n_1 - n_2$ varies with temperature, thermal stabilization is required and it is actively enforced by a PI loop. The NTT device integrates both a thermistor to sense the crystal temperature and a Peltier cell. An home-made PI loop stabilizes the thermistor reading to within a few mK,  
independently on the fundamental power injected, thus the temperature instability is well below the width of the peak in the SHG efficiency  versus temperature, equal to 2.4~K (FWHM), as shown in \fref{fig:SHGeff}(a). 

Also, $n_1 -n_2$ depends on the wavelength, thus quasi phase-matching temperature $T_Q$ 
changes with the fundamental wavelength $\lambda_1$, at a measured rate of $d T_Q/d\lambda_1 = -11$~K/nm. It follows that, at a given temperature, SHG filters the ASE in the fundamental power off a spectral window $ \Delta \lambda_1 \simeq 2.4/11 \simeq 0.2 $~nm.

At the peak temperature, we measured the SHG efficiency by varying the EDFA  output power $P_{1560}$ up to 25~dBm, {\it i.e.} approximately 0.32~W, where we reached a SH power $P_{780}$ of 93~mW. This amounts to a conversion efficiency $P_{780}/P_{1560} = 0.29$, which corresponds to a non-linear efficiency $\eta = P_{780}/P_{1560}^2 = 0.92$~W$^{-1}$. When the converted SH power is negligible with respect to input power ("undepleted pump regime"), we expect $\eta$ to be independent of $P_{1560}$. \fref{fig:SHGeff}(b) shows that this is the case only for input power below $\sim 50$~mW, where we measured the non-linear conversion of approximately 2~W$^{-1}$ in agreement with the manufacturer specifications. In a previous set of measurements, we had increased the EDFA power to 0.5~W and obtained 0.27~W of SH at 780~nm. This level of power is perfectly adequate for a laser cooling experiment with Rb atoms, especially as we take into account that the light is output from a fiber, thus with a clean spatial mode. As a rule of thumb, for the light amplified by semiconductor tapered amplifiers (TA), which are commonplace in laser cooling experiments, the fiber injection efficiency is of the order of 50\%.

\section{Frequency noise and amplitude stability at 780~nm}

Phase noise on the 1560nm-amplification and the SHG could increase the linewidth of the lasers, therefore it is advisable to repeat the heterodyne measurement also at 780~nm. In addition we measure the long-term amplitude and polarization stabilities. All measurements were performed with the EDFA set at 27dBm power output.

\subsection{Linewidth at 780~nm}

\begin{figure}[t]
\centering
\begin{tikzpicture}
\node at (0,0) {    
\includegraphics[width=0.475\textwidth]{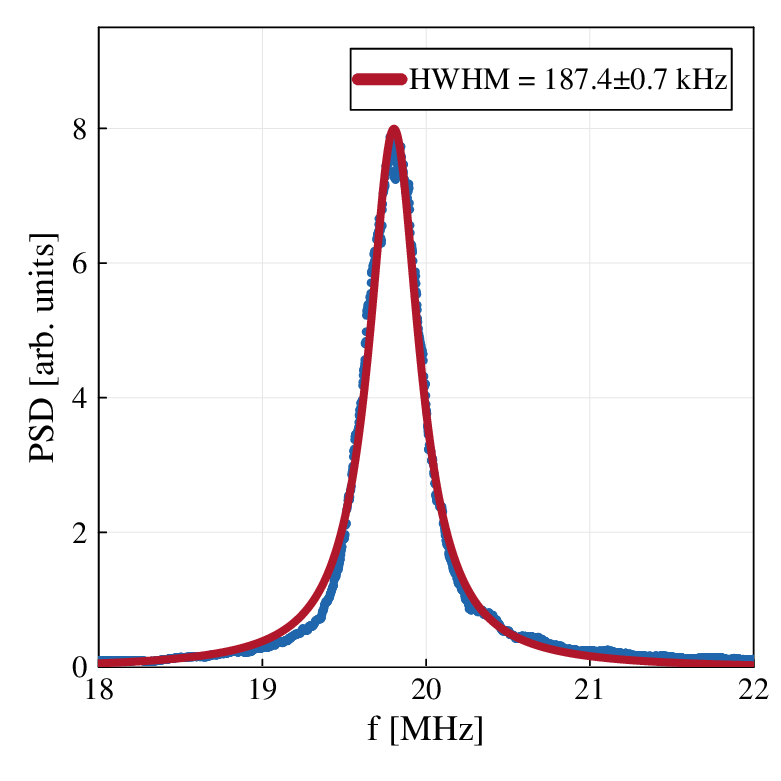}};
\node at (6.75,0) {
\includegraphics[width=0.475\textwidth]{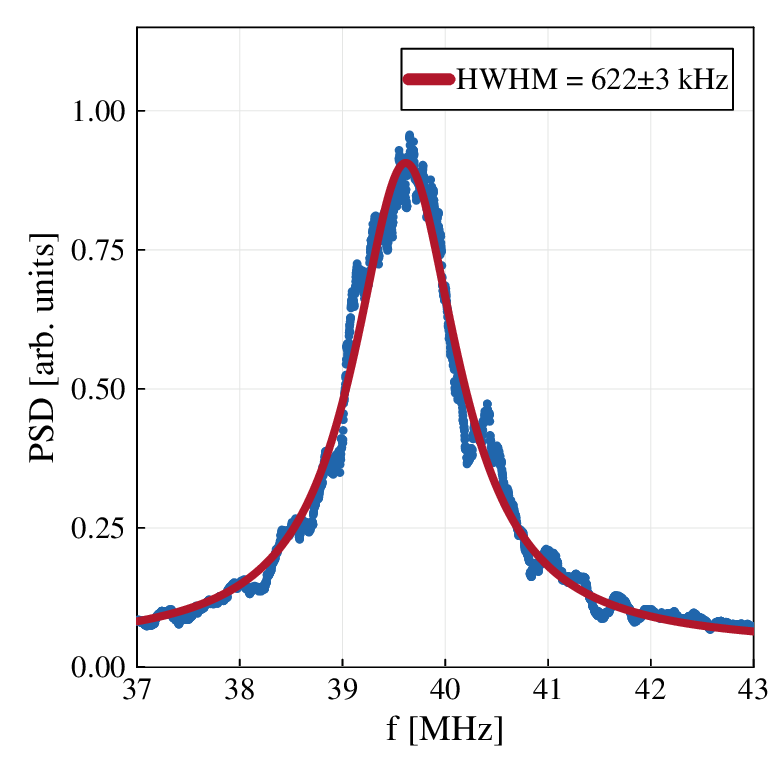}};
\node at (-3,2.75) {\LARGE{(a)}};
\node at (3.75,2.75) {\LARGE{(b)}};
\end{tikzpicture}
\caption{PSD of the beat notes calculated as the FFT of the digital scope time traces at: (a) $\Delta_{12}/(2\pi) \simeq  20$~MHz with half-width at half-maximum equal to 187.4~kHz; (b) $2\Delta_{12}/(2\pi) \simeq 40$~MHz, with half-width at half-maximum equal to 622~kHz. A moving average over 101 point is applied upon the $y$ vector of 9001 points; units are the same in the two panels.}
\label{fig:beat780}
\end{figure}

To observe the beatnote after SHG, the InGaAS photodiode has been replaced with a Si photodetector (Hamamatsu G4176-03). The detector is biased with a 9V battery through a bias-tee (Minicircuits ZX85-12G-S+) that allows to feed the dc bias and extract the ac signal. 

It must be noted that when the two fundamental frequencies are as close as $\Delta_{12} = 2\pi\times 20$~MHz, at 780~nm we observe not only the beatnote between the two frequency-doubled sources at $2\Delta_{12}$ but also a beatnote, actually stronger, at $\Delta_{12}$, as shown in \fref{fig:beat780}. Alongside SHG, sum-frequency generation (SFG) takes place, whereby two fields $(\omega_1, \omega_2)$ generate a field at the sum frequency $\omega_1+\omega_2$. The beatnote of the latter field with any of the two SHG beams generates a signal at frequency $\Delta_{12}$. Indeed the complex amplitude of the field at 780~nm emitted by the non-linear crystal of the length $L$, $E_3(L,t)$, contains three frequency components \cite{Boyd:2012}:

\begin{align}
    E_3(L,t) &= \frac{i\epsilon_0 \mu_0 \chi^{NL} L}{2 k_3} 
    \left[
    4 \omega_1^2 |E_1|^2 e^{-i(2\omega_1 t + 2\phi_1)} +
    4 \omega_2^2 |E_2|^2 e^{-i(2\omega_2 t + 2\phi_2)} + \right. \nonumber \\
    &
    \quad + \left. 2 (\omega_1+\omega_2)^2 |E_1 E_2| e^{-i(\omega_1 t + \omega_2 t + \phi_1 + \phi_2)} 
    \right]
\end{align}
where $\chi^{NL}$ is the relevant non-linear susceptibility and perfect quasi-phase matching is assumed for all conversions. 

As noted above in Eq.~\ref{eq:PSD}, the PSD of the 780nm-detector photocurrent is proportional to the Fourier transform of the 780nm-light intensity auto-correlation function

\begin{align}
        R(\tau ) &\equiv \langle E_3^*(t)E_3^*(t+\tau)E_3(t+\tau)E_3(t) \rangle \\
         & \propto  (|E_1|^4 + |E_2|^4 + 4 |E_1|^2|E_2|^2)^2 + \nonumber \\
    & \quad + |E_1|^4|E_2|^4 e^{-2\langle \varphi_{12}^2(\tau) \rangle}
    \left( e^{i 2 \Delta_{12}\tau } + \mathrm{c.c.} \right) + \nonumber \\
    & \quad + 4|E_1|^2|E_2|^2 (|E_1|^2+|E_2|^2)^2 e^{-\frac12\langle \varphi_{12}^2 (\tau) \rangle}
    \left( e^{i \Delta_{12}\tau } + \mathrm{c.c.} \right).
\end{align}

Thus the $PSD_i$ component at $\omega=\Delta_{12}$ is expected to display the same broadening as the beatnote of the two input fields at 1560~nm, since both are due to the same phase noise factor 
$\exp [-\langle \varphi_{12}^2 (\tau) \rangle/2]$. 
From \fref{fig:beat780}(a) we observe that the FWHM here is approximately the same as seen above, in Subsection \ref{subsec:freq_noise1560} and \fref{fig:beat1560}, thus we conclude that the fields at SHG input have nearly the same frequency spectrum of the two seed sources, with no appreciable impact of the EDFA amplification.

As mentioned above, even for the beatnote at $\omega = 2\Delta_{12}$ the FWHM depends on the type of phase noise.  In case of white frequency noise, the FWHM of the component at $\omega=2\Delta_{12}$ is four times larger than the component at $\omega=\Delta_{12}$. This is in qualitative agreement with the observed FWHM$_{780}/$FWHM$_{1560}$~=~3.4.

We conclude that each beam at 780~nm has a linewidth of approximately 0.6~MHz, determined by the linewidth of the corresponding seed laser at the fundamental wavelength of 1560~nm.

\subsection{Power and polarization stability}

\begin{figure}[t]
\centering
\begin{tikzpicture}
\node at (0,0) {    
\includegraphics[width=0.475\textwidth]{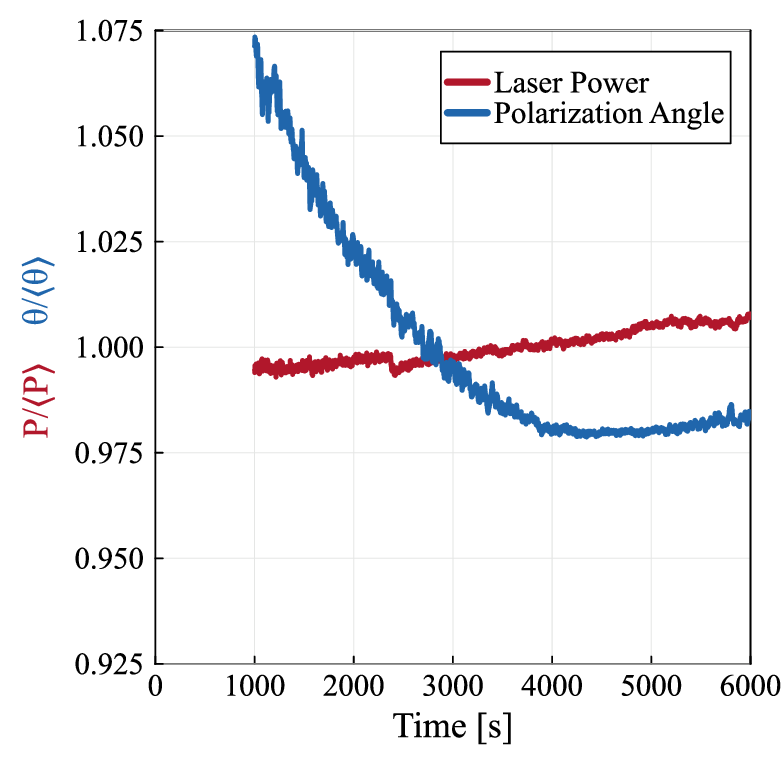}};
\node at (6.75,0) {
\includegraphics[width=0.475\textwidth]{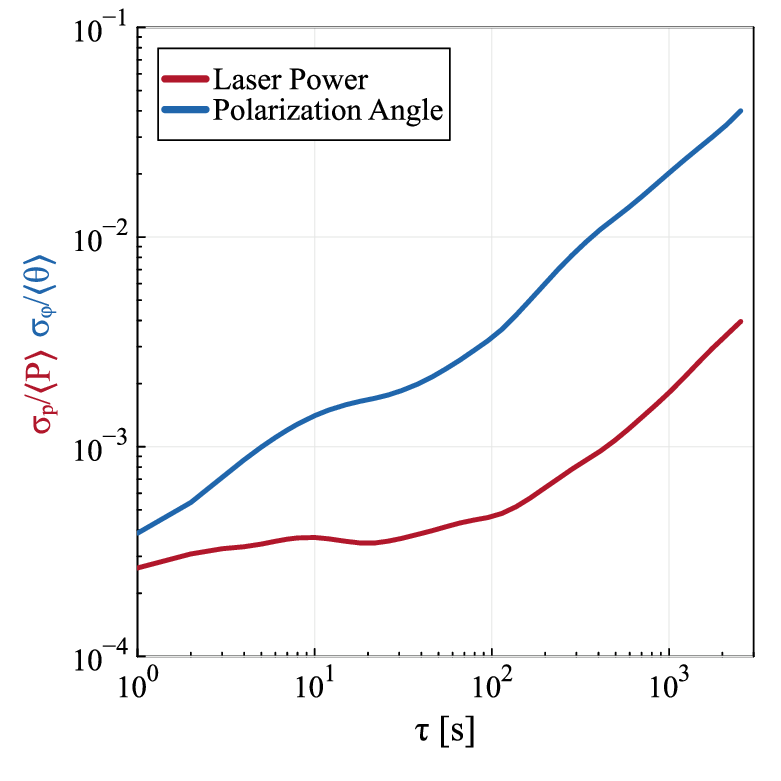}};
\node at (-3,2.75) {\LARGE{(a)}};
\node at (3.75,2.75) {\LARGE{(b)}};
\end{tikzpicture}
\caption{(a) SHG power (red) and polarization angle (blue) as a function of time, normalized to their average value, after a 1000~s warm-up period; (b) relative Allan deviations calculated from the acquired data.}
\label{fig:stability780}
\end{figure}

Finally, we have measured the long-term stability of the SH power at 780~nm simply by reading the photodetector (Thorlabs PM400K3) with a sample rate of 1~s for approximately 3~hours. To assess any potential instability of polarization we have repeated the same measurement with the addition of a polarizer inserted between the fiber output and the photodetector. 

After removing the data of the initial 1000~s warm-up, we notice that the power is stable within $\sim 1\%$ [see \fref{fig:stability780}(a)]. To be more quantitative, we show the fractional Allan deviations of the power and the polarization angle [see \fref{fig:stability780}(b)]: in the long term, both increase with integration time $\tau$, as $\sigma_{y} \sim \tau^{0.6}$.

\section{Conclusions}
We have designed, implemented and characterized a laser system suitable for laser cooling of Rb atoms. Our cost-effective system \cite{cost_note} is entirely realized with fiber components to ensure compactness and robustness, and to avoid the need of optical alignments. With the exception of the control electronics, we have employed all commercial components, taking advantage of the many available in the telecom spectral region around 1550~nm. 

Two low-power semiconductor lasers at 1560~nm are combined and fed to a single commercial EDFA module. The amplified radiation undergoes SHG to 780~nm in a PPLN crystal. As a side effect, also SFG takes place and is significant. This represents a potential loss of power; even more importantly, this populates the spectrum of beatnotes observed at 780~nm and complicates the frequency stabilization of the sources (not described), achieved through the beat note signals with a third laser (reference) at 780~nm.

We have obtained up to 300 mW at 780~nm with a linewidth of 0.6~MHz, limited by the linewidth of the laser sources at 1560~nm, which is perfectly adequate for laser cooling of Rb, for which the D2 transition has a natural linewidth 10~times larger. Indeed our system, frequency stabilized to a low-power 780~nm reference laser (itself frequency locked to Rb saturated absorption), is ready to be employed for a magneto-optical trap.

\begin{backmatter}
\bmsection{Funding}
Funding is acknowledged through the project CRYST$^3$, under the European Union’s Horizon 2020 Research and Innovation programme, Grant Agreement No. 964531.

\bmsection{Acknowledgments}
We are grateful to Leonardo Razzai for his help in gathering and analyzing the beatnotes and stability data.

\bmsection{Disclosures} The authors declare no conflicts of interest.

\bmsection{Data availability} Data underlying the results presented in this paper are available in \cite{dataset}.


\end{backmatter}

\bibliography{refs}

\begin{thebibliography}{10}
\bibitem{TIA_input_noise}
We neglect the TIA input current noise, with a nominal value of 0.13 pA/$\sqrt{\si{\hertz}}$.
\bibitem{cost_note}
The total cost of the described components, not including home-made electronics, was approximately 40k Euro.
\end{thebibliography}

\end{document}